\documentclass[]{spie}  

\usepackage{amsmath,amsfonts,amssymb}
\usepackage{graphicx}
\usepackage[colorlinks=true, allcolors=blue]{hyperref}

\title{The multi-physics analysis, design and testing of CUSP, a CubeSat mission for space weather and solar flares x-ray polarimetry}

\author[a,c]{Giovanni~Lombardi}
\author[a]{Sergio~Fabiani}
\author[a]{Ettore~Del~Monte}
\author[a,b]{Andrea~Alimenti}
\author[g,m]{Riccardo~Campana}
\author[h]{Mauro~Centrone}
\author[a]{Enrico~Costa}
\author[a]{Nicolas~De~Angelis}
\author[g]{Giovanni~De~Cesare}
\author[a]{Sergio~Di~Cosimo}
\author[a]{Giuseppe~Di~Persio}
\author[a]{Abhay~Kumar}
\author[a]{Alessandro~Lacerenza}
\author[a]{Pasqualino~Loffredo}
\author[k]{Gabriele~Minervini}
\author[a]{Fabio~Muleri}
\author[l]{Paolo~Romano}
\author[a]{Alda~Rubini}
\author[a]{Emanuele~Scalise}
\author[a,b]{Enrico~Silva}
\author[a]{Paolo~Soffitta}
\author[d]{Davide~Albanesi}
\author[e]{Ilaria~Baffo}
\author[f]{Daniele~Brienza}
\author[d]{Valerio~Campamaggiore}
\author[i]{Giovanni~Cucinella}
\author[j]{Andrea~Curatolo}
\author[d]{Giulia~de~Iulis}
\author[d]{Andrea~Del~Re}
\author[i]{Vito~Di~Bari}
\author[i]{Simone~Di~Filippo}
\author[f]{Immacolata~Donnarumma}
\author[e]{Pierluigi~Fanelli}
\author[j]{Nicolas~Gagliardi}
\author[d]{Paolo~Leonetti}
\author[f]{Matteo~Mergè}
\author[j,k]{Dario~Modenini}
\author[i]{Andrea~Negri}
\author[j]{Daniele~Pecorella}
\author[i]{Massimo~Perelli}
\author[k]{Alice~Ponti}
\author[d]{Francesca~Sbop}
\author[j,k]{Paolo~Tortora}
\author[f]{Alessandro~Turchi}
\author[f]{Valerio~Vagelli}
\author[f]{Emanuele~Zaccagnino}
\author[d]{Alessandro~Zambardi}
\author[e]{Costantino~Zazza}

\affil[a]{INAF-IAPS\\ via del Fosso del Cavaliere 100, 00133, Rome, Italy}
\affil[b]{Department of Industrial, Electronic and Mechanical Engineering, "Roma Tre" University, via V. Volterra 62, 00146, Rome, Italy}
\affil[c]{Department of Enterprise Engineering ”Mario Lucenti”, University of Rome "Tor Vergata", Via Cracovia 50, 00133, Rome, Italy}
\affil[d]{DEDA Connect s.r.l.\\ via Vincenzo Lamaro 51, 00173 Rome, Italy}
\affil[e]{DEIM, University of "La Tuscia", Largo dell’Università, 01100, Viterbo, Italy}
\affil[f]{ASI, via del Politecnico snc\\ 00133, Rome, Italy}
\affil[g]{INAF-OAS Bologna\\ via Gobetti 93/3, 40129, Bologna, Italy}
\affil[h]{INAF-OAR\\ via Frascati 33, 00040, Monte Porzio Catone, Italy}
\affil[i]{IMT s.r.l.\\ via Carlo Bartolomeo Piazza 30, 00161, Rome, Italy}
\affil[j]{Alma Mater Studiorum University of Bologna - Department of Industrial Engineering and Interdepartmental Center for Industrial Aerospace Research, Via Fontanelle 40, 47121, Forl\`i, Italy}
\affil[k]{INAF-Headquarters\\ viale del Parco Mellini 84, 00136, Rome, Italy}
\affil[l]{INAF-OACT\\ Via S. Sofia 78, 95123, Catania, Italy}
\affil[m]{INFN-Bologna, viale Berti Pichat 6/2,  40127, Bologna, Italy}

\authorinfo{Further author information: (Send correspondence to G. Lombardi)\\ E-mail: giovanni.lombardi@inaf.it, Telephone: +39 3458583819}

\begin{document} 
\maketitle

\begin{abstract}
The space$-$based CUbesat Solar Polarimeter (CUSP) mission aims to measure the linear polarization of solar flares in the hard X-ray band by means of a Compton scattering polarimeter. CUSP is a project in the framework of the Alcor Program of the Italian Space Agency aimed at developing new CubeSat missions. As part of CUSP’s Phase B study, which began in December 2024 and will last one year, we present the current development status of the design solutions adopted for the mission’s most critical multi$-$physics design drivers. These solutions have been formulated and applied to demonstrate compliance with system requirements at both the spacecraft and platform levels. In particular, we describe the mechanical design of each structural component, the results of static, dynamic finite element analyses, and a proposal for topological optimization of the interface between the platform and payload and some fixture for test, and the preliminary environmental testing campaign (e.g., vibration, shock) that will be carried out on a mechanical demonstrator.  
\end{abstract}


\section{INTRODUCTION}
\label{sec:intro}  
The CUbesat Solar Polarimeter (CUSP)\cite{fabiani1} is a CubeSat mission in low Earth orbit, specifically conceived to measure the linear polarization of solar flares in the hard X-ray range (25--100 keV) through a Compton scattering polarimeter. This cutting-edge mission is expected to provide new insights into magnetic reconnection and particle acceleration mechanisms occurring within the Sun’s flaring magnetic structures, which are crucial processes underlying space weather dynamics.\\
Hard X-ray polarimetry offers a unique observational window into the physical mechanisms responsible for particle acceleration, as well as the temporal evolution of magnetic configurations during solar flares. These data are essential for refining current models of solar activity and understanding its influence on the heliosphere. Such improved knowledge will contribute to the development of more accurate space weather predictions, which are essential to mitigate the risks for satellite systems, communication infrastructures, and terrestrial power grids.\\
CUSP is developed within the framework of the Alcor Program, a strategic initiative led by the Italian Space Agency (ASI) to promote innovative CubeSat missions. Having been approved for Phase B in 2024\cite{fabiani2024cusp}, the project is currently advancing through a series of multiphysics analyses and engineering validations focused on mechanical, thermal, and structural integrity.\\
In particular, extensive finite element analyses (FEA) have been carried out to assess the payload's behavior under launch and in-orbit environmental conditions. The structural model includes detailed representations of all major mechanical interfaces and subassemblies, taking into account realistic constraints and contact conditions.\\ These simulations are being used to predict natural frequencies, stress distributions, and dynamic response to sine and random vibrations, in compliance with ECSS\cite{ecss} and GEVS\cite{gevs} standards. The objective is to ensure that the payload can withstand the mechanical loads encountered during critical mission phases without failure.\\
The outcomes of the numerical analyses have informed the optimization of mechanical fixtures for environmental testing and will serve as the baseline for upcoming qualification tests on a vibration shaker. These tests, planned for the next development phase, are crucial for correlating numerical models with physical behavior and for de-risking the integration of the payload into the CubeSat platform.\\
This paper presents the preliminary design solutions proposed for the CUSP payload, with a focus on the multiphysics simulations performed and the engineering strategies adopted to validate the mechanical design prior to flight.
The paper is structured as follows: Section Workflow and Software describes the design workflow and software tools adopted; Section Payload Design focuses on the payload design strategy; Section Multi-Physics analyzes and Results presents the results of the multiphysics simulations; and finally, Section Conclusions summarizes the conclusions and outlines future activities.

\section{WORKFLOW AND MODELING STRATEGY}
The multifaceted design of a CubeSat mission introduces a wide range of engineering challenges. These include the necessity to work within strict volumetric constraints, adhere to severe mass limitations for each subsystem, and comply with rigorous scientific and mechanical requirements. In response to these demands, a mechanical baseline was established and continuously refined for the core of the instrument, the sensitive heart of CUSP, while balancing the interplay of the mission’s multidisciplinary constraints.\\
Starting from the conceptual design developed during Phase A, a systematic approach was adopted to assess each mechanical component and its interfaces—both internal and external—to the payload. Special attention was paid to the correct modeling of contact and constraint conditions to ensure fidelity to the real operational setup.\\
An iterative optimization procedure was subsequently implemented, integrating mechanical, thermal, and optical aspects to support informed design decisions. The methodology focused on progressively enhancing the model accuracy while preserving numerical efficiency for simulation purposes.\\
The design and analysis activities relied on the following software tools:
\begin{itemize}
\item \textbf{SolidWorks}: Used to develop the full parametric 3D CAD model of the payload structure, facilitating design iterations and interface definition\cite{dass}.
\item \textbf{ANSYS SpaceClaim}: Employed for model defeaturing and geometric preparation, enabling the removal of non-structural features such as fillets and secondary holes.
\item \textbf{ANSYS Meshing}: Responsible for generating high-quality structured and unstructured meshes, with refinement in regions of interest.
\item \textbf{ANSYS Workbench}: Served as the integrated simulation environment, managing the entire workflow from preprocessing through postprocessing, and enabling multiphysics coupling and parametric studies\cite{ansys}.
\end{itemize}

\begin{figure}[h!]
\centering
\includegraphics[scale=0.5]{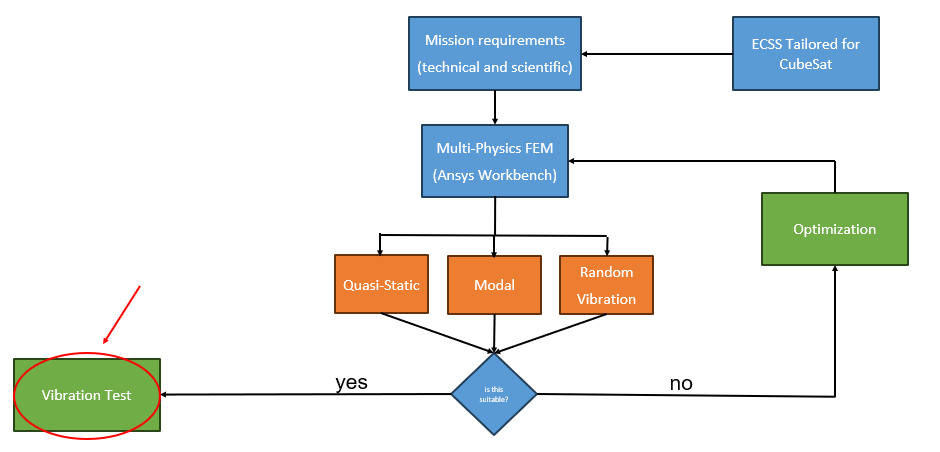}
\caption{Workflow for method analysis}
\label{workflow}
\end{figure}

The modeling process followed a sequential strategy:
\begin{itemize}
\item In-depth examination of the mechanical architecture and definition of subsystem roles and load paths.
\item Simplification of the CAD model to retain only structurally relevant features, ensuring that mass and stiffness characteristics were preserved.
\item Definition and application of appropriate boundary conditions and loads representing the launcher and orbital environments.
\item Execution of finite element simulations to evaluate modal behavior, stress distribution, and structural margins in compliance with ECSS and GEVS standards.
\end{itemize}

\newpage

\section{PAYLOAD DESIGN}
\label{sec:sections}
\begin{figure}[h!]
\centering
\includegraphics[scale=0.5]{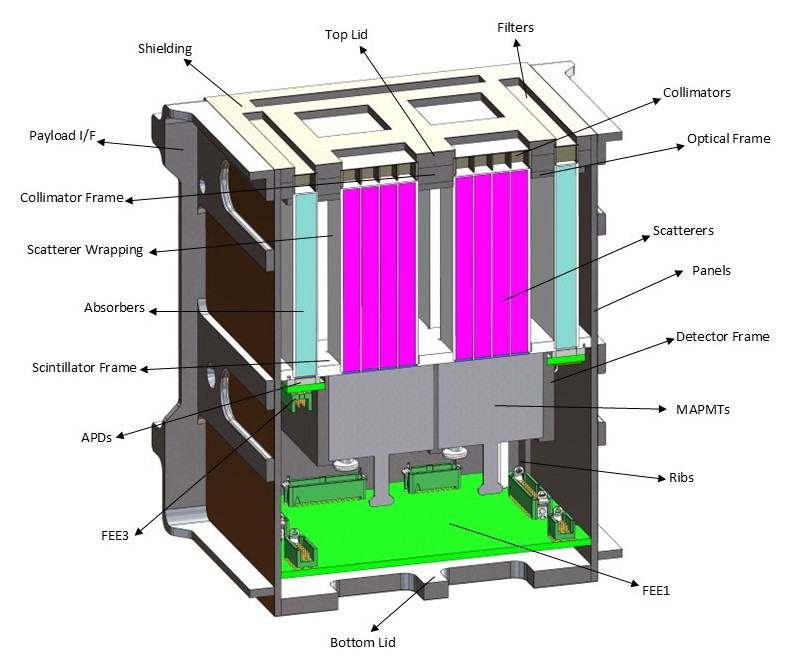} 
\caption{Scheme of CUSP payload} 
\label{figure2} 
\end{figure}
\begin{figure}[h!]
        \centering
        \includegraphics[scale=0.35]{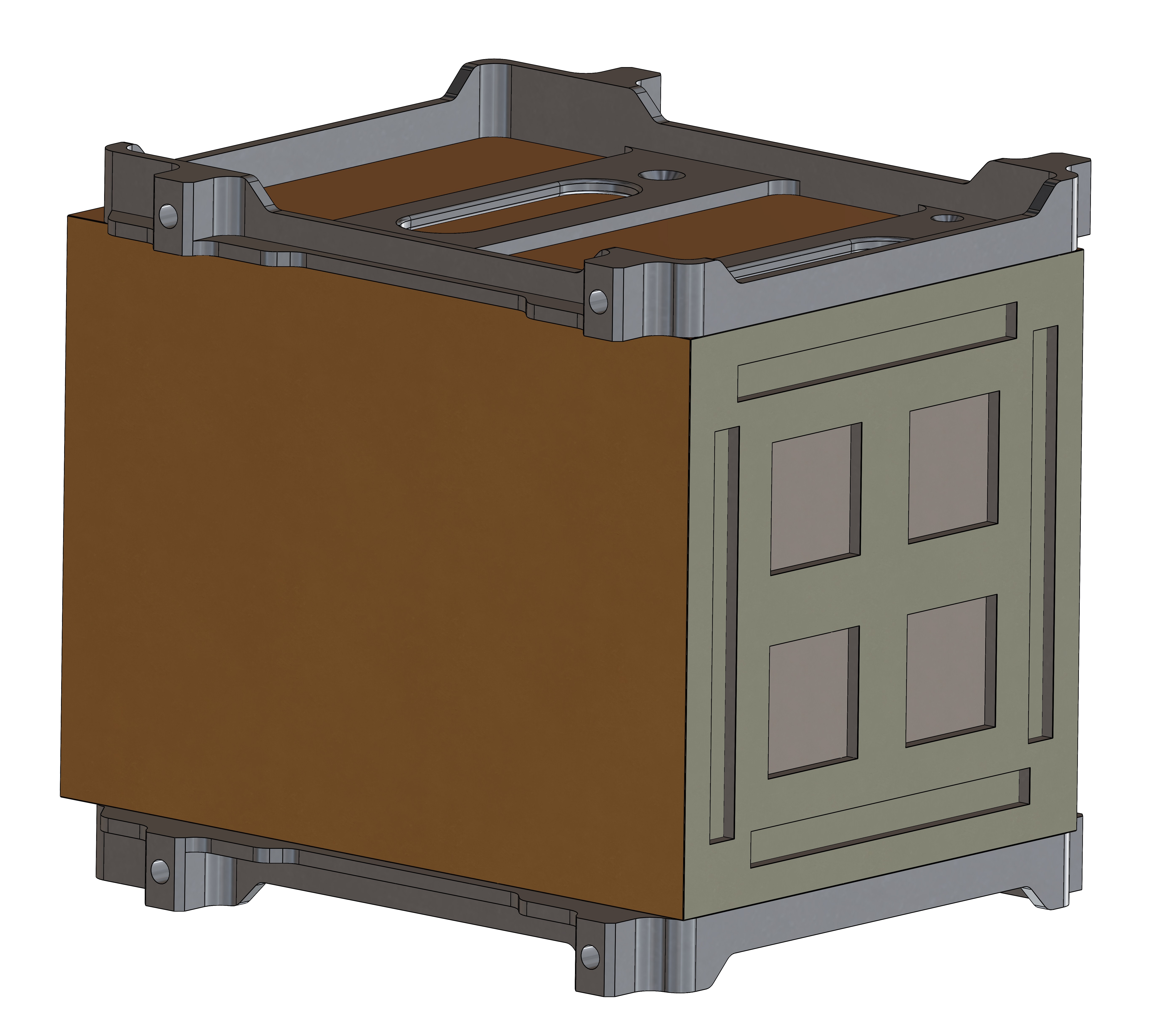}\quad\\
        \includegraphics[scale=0.7]{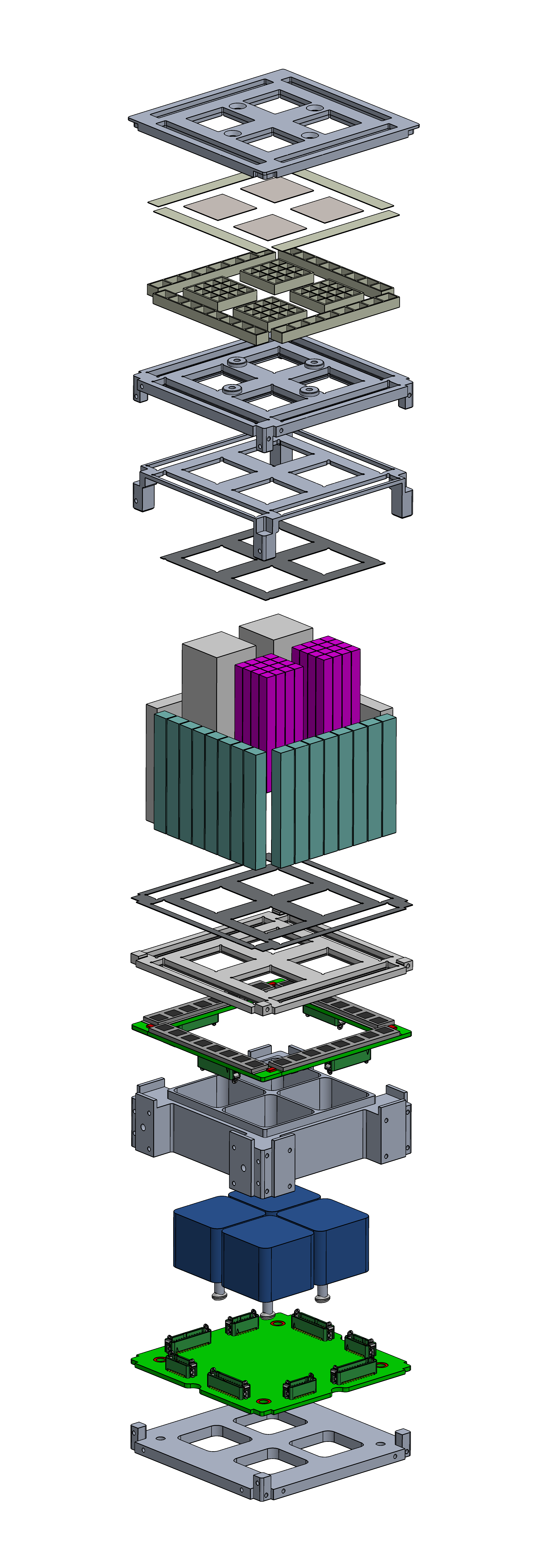}
        \caption{Payload main unit and exploded view}
        \label{figure34}
\end{figure}

 
The mechanical configuration of the CUSP payload, as shown in Figure~\ref{figure2}, has been iteratively refined to address the strict constraints typical of CubeSat architectures. The structural layout is primarily composed of high-strength, lightweight aluminum alloy components, precisely arranged to endure both mechanical and thermal loads while integrating scientific instrumentation and electronics.\\
A key development in the design was the introduction of a novel structural-optical element, termed the \textit{optical frame}. This component was engineered to provide precise alignment and rigid mechanical support for the plastic scintillators and GAGG crystal arrays. Considering the payload's sensitivity to misalignments, the optical frame was designed to maintain geometrical stability across all mission phases, including the high dynamic loading conditions during launch.\\
Further refinements were implemented on the detector frame, which was adapted to enable the integration of the third Front-End Electronics board (FEE3). These updates required careful modifications of the frame geometry and relative positioning, ensuring compliance with mechanical tolerances and maintaining effective thermal pathways to preserve the structural and thermal integrity of the system.\\
Figure~\ref{figure34} illustrates the main unit of the payload and its exploded view. The core subsystems include: 
\begin{itemize}
 \item {Two identical aluminum brackets for mechanical interface with the satellite platform;}
 \item {A top and bottom enclosure in aluminum alloy, plus four side panels with passive tungsten radiation shielding;}
 \item {A machined collimator tray holding four collimators with filters for scatterers for absorbers;}   
\item {32 absorber modules made of GAGG scintillators, each mounted on a PCB with integrated APD readout;}   
\item {64 scatterer elements made of plastic scintillator, supported within the same mechanical and electrical structure as the absorber tray, and mounted on a dedicated support frame;}
\item {One FEE1 board interfacing with the BEE (Back-End Electronics) in the secondary instrument unit;}
\item {FEE boards for signal processing of MAPMTs and APD arrays;}
\item { Four structural ribs to ensure stiffness, load distribution, and internal alignment.}
\end{itemize}

\section{MULTI-PHYSICS FEM AND RESULTS}
\begin{figure} [ht]
\centering
\includegraphics[scale=0.35]{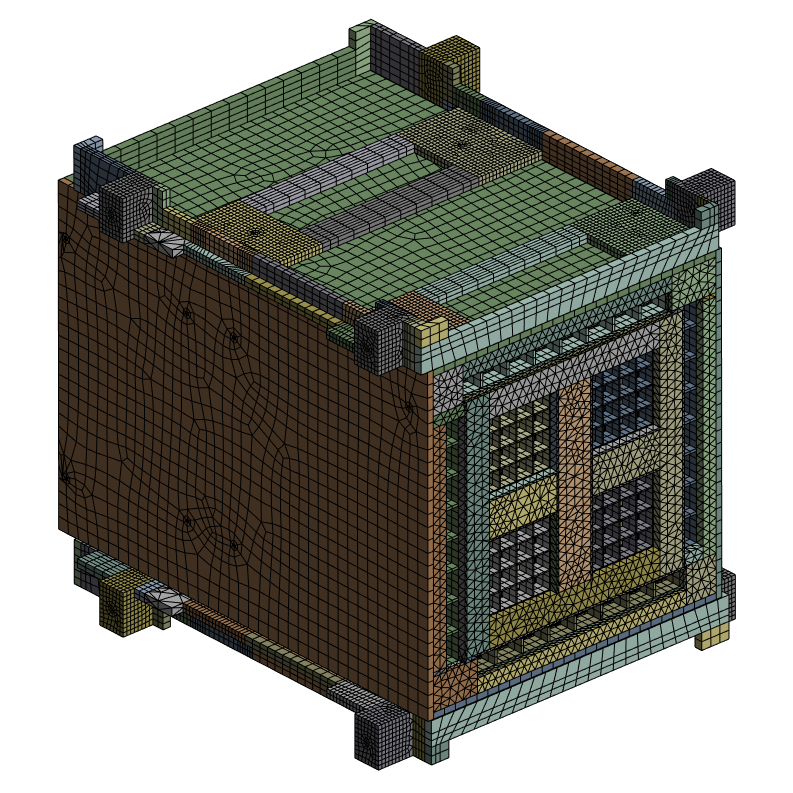}
\caption{Detailed mesh of the payload main unit}
\label{figure5}
\end{figure}

\begin{figure}[h!]
    \centering
    \includegraphics[scale=0.4]{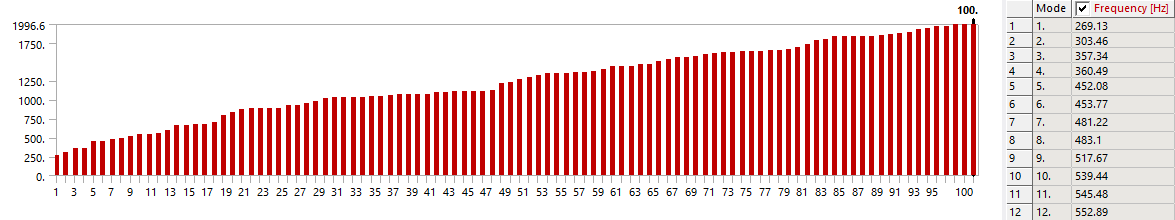}
    \caption{Results from modal analysis. The first six modes represent rigid body motion; elastic modes follow.}
    \label{figure6}
\end{figure}

\begin{figure}[h!]
        \centering
        \includegraphics[scale=0.3]{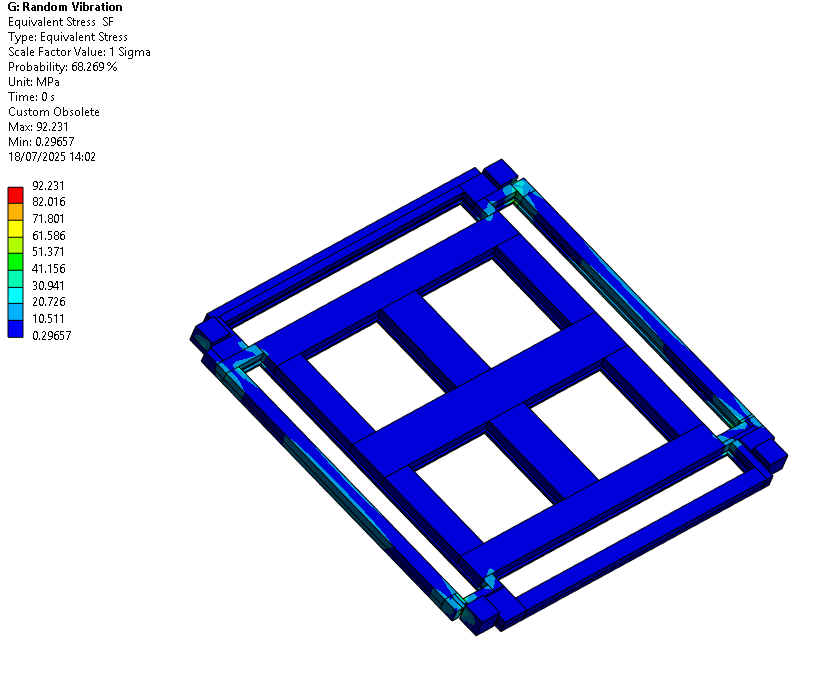}\quad
        \includegraphics[scale=0.3]{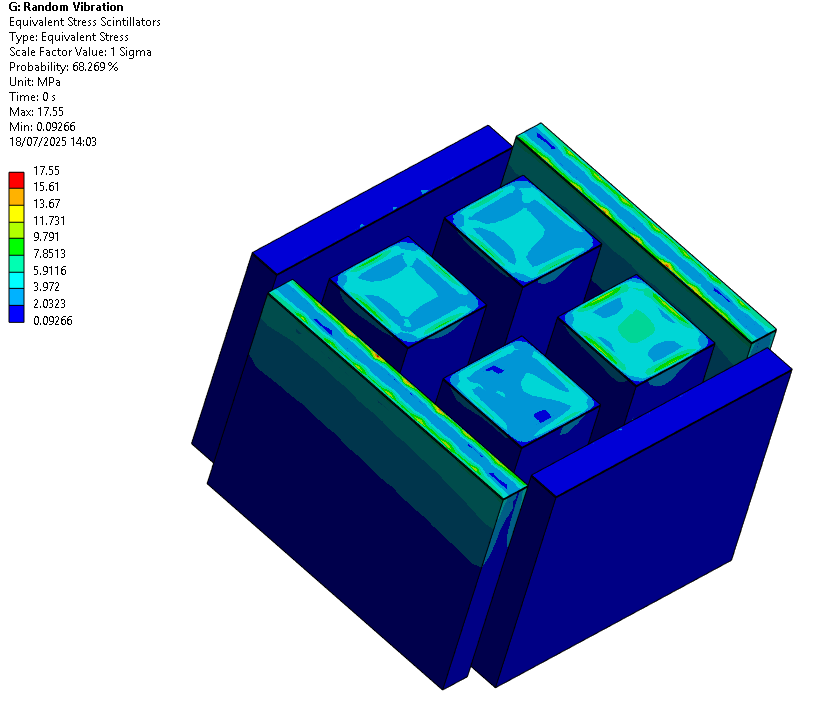}
        \caption{Stress response from random vibration analysis.}
        \label{figure8}
\end{figure}

\begin{figure}[h!]
        \centering
        \includegraphics[scale=0.20]{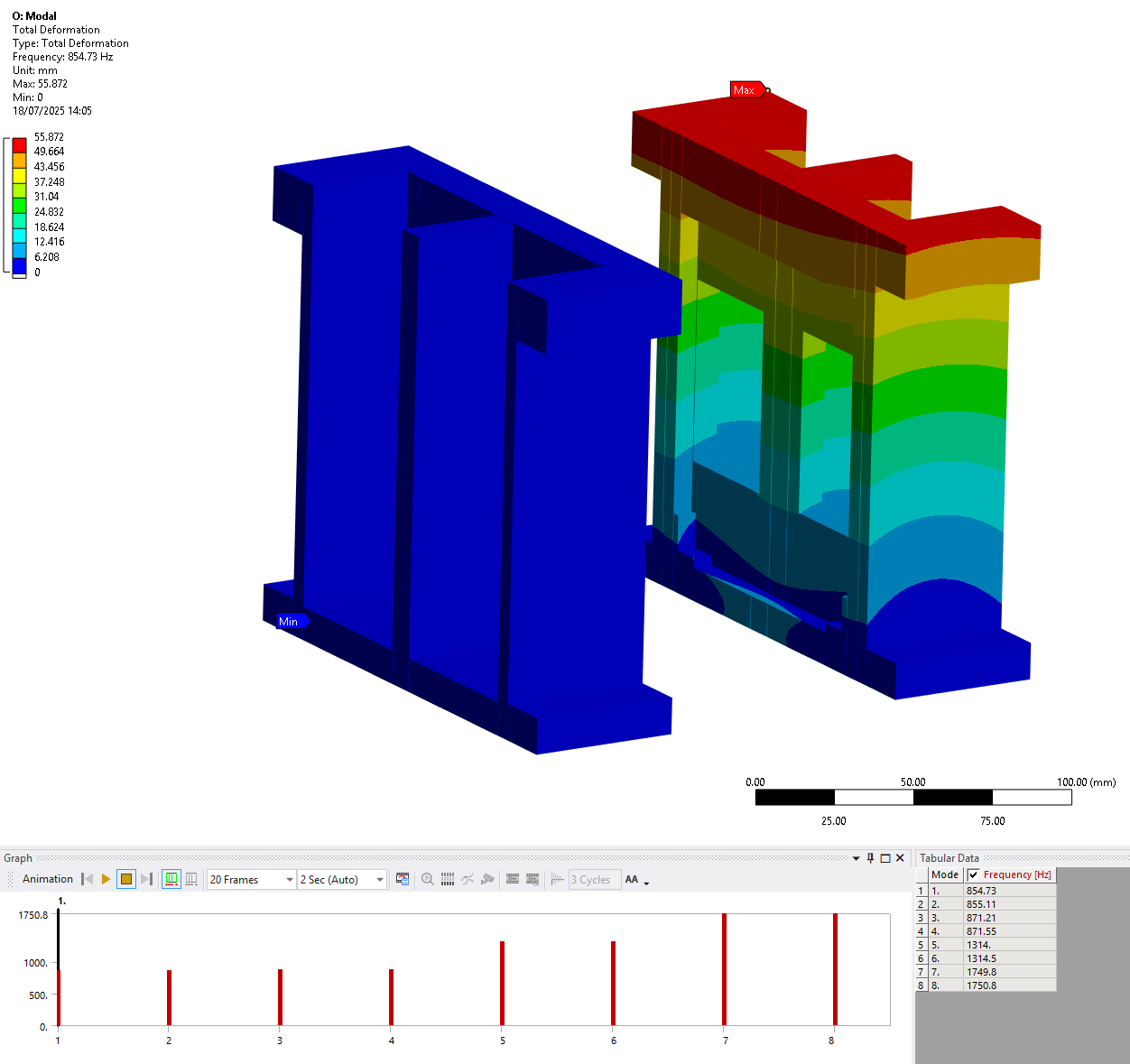}\quad
        \includegraphics[scale=0.20]{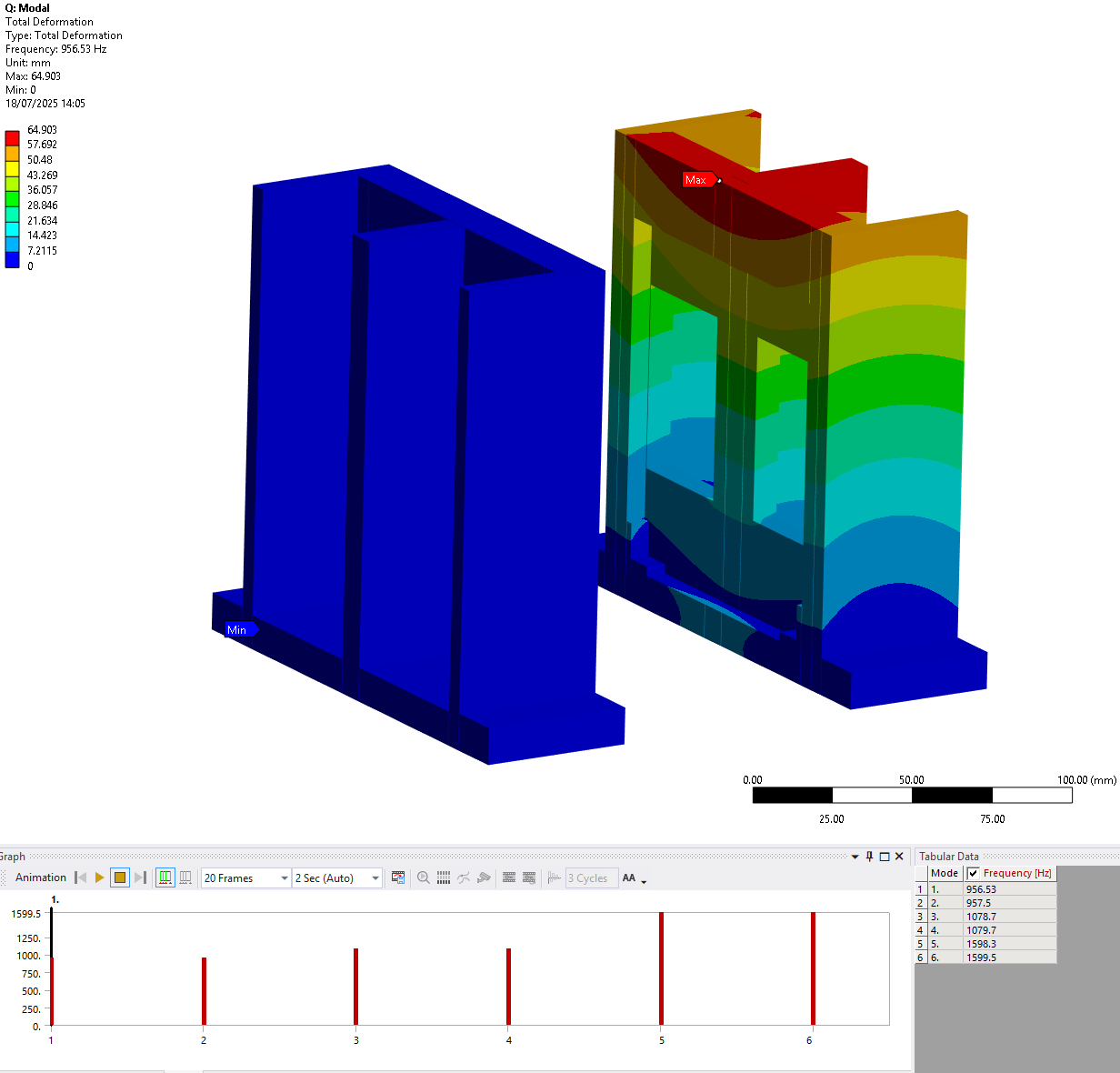}
        \caption{Optimized test fixture for environmental qualification.}
        \label{figure10}
\end{figure}

The multiphysics simulation campaign for the CUSP payload was carried out using the ANSYS Workbench suite, leveraging its advanced capabilities for structural and thermal analysis. A simplified yet representative finite element model was developed, preserving the global envelope and mass distribution while reducing geometrical complexity to minimize computational overhead. Special emphasis was placed on maintaining accurate representations of interfaces and mechanical constraints to ensure model fidelity.\\
A high-fidelity mesh\cite{schultz2018} was generated, as shown in Figure~\ref{figure5}, with local refinements in critical regions to accurately capture stress concentrations and deformation gradients. Such detailed meshing is indispensable for resolving complex physical responses, particularly those involving small-scale features.
Several components, such as the scintillator blocks wrapped in epoxy and mechanically housed via interference fits, are not rigidly bonded to the surrounding structures. To accurately replicate their behavior under load, contact definitions using frictionless or weakly coupled conditions\cite{johnson2004} were applied. These approaches avoid artificial stiffening effects that could result from overly restrictive contact assumptions. Screw and bolt interfaces were modeled through remote point constraints combined with preloaded contact definitions.\\
This methodology ensures accurate transmission of mechanical loads and realistic simulation of bolted joint behavior under dynamic excitation. Each structural component was assigned material properties either from the ANSYS material library or from customized definitions based on experimental data and datasheets.
\begin{table}[ht]
\caption{Material properties} 
\label{tab1}
\begin{center}       
\begin{tabular}{|l|l|l|l|l|l|}
\hline
\rule[-1ex]{0pt}{1ex}  \textbf{Material} & \textbf{$\rho$ (kg/m$^3$)} & \textbf{E (GPa)} & \textbf{$\nu$} & \textbf{YS (MPa)} & \textbf{US (MPa)} \\
\hline
\rule[-1ex]{0pt}{1ex}  Aluminum 7075 & 2810 & 71.7 & 0.33 & 503 & 572 \\
\hline
\rule[-1ex]{0pt}{1ex}  Resin Epoxy & 1160 & 37 & 0.35 & 54.6 & 54.6 \\
\hline
\rule[-1ex]{0pt}{1ex}  FR4 & 1944 & 24 & 0.16 & 298 & 298 \\
\hline
\rule[-1ex]{0pt}{1ex}  Ti6Al4V & 4430 & 113.8 & 0.34 & 880 & 950 \\
\hline 
\rule[-1ex]{0pt}{1ex}  PEEK & 1320 & 3.6 & 0.36 & 90 & 100 \\
\hline 
\rule[-1ex]{0pt}{1ex}  Tungsten & 19300 & 400 & 0.28 & 550 & 750 \\
\hline 
\rule[-1ex]{0pt}{1ex}  Scintillator GAGG & 6630 & 1.86 & 0.4 & 28 & 29.2 \\
\hline 
\end{tabular}
\end{center}
\end{table}
A series of finite element analyses were carried out to evaluate the payload’s mechanical robustness and its compliance with space qualification standards. These included:
\begin{itemize}
\item \textbf{Quasi-static analysis}, with load factors of 20g applied independently along each principal axis to simulate the structural response to launch-induced accelerations.
\item \textbf{Random vibration analysis}, using an interpolated Power Spectral Density (PSD) curve derived from ESA ECSS and NASA GEVS standards, to validate the payload against vibrational fatigue.
\item \textbf{Modal analysis}, confirming that the first natural frequency exceeds 120 Hz\cite{holmes2010}, thus satisfying the required margin above the 100 Hz lower limit specified in the ECSS guidelines.
\end{itemize}
Together, these simulations constitute a comprehensive verification campaign, providing robust evidence for the mechanical integrity and environmental readiness of the CUSP payload ahead of upcoming qualification tests on a shaker system.\\
The structural verification activities performed on the CUSP payload yielded highly promising results, confirming the design's compliance with the mission's mechanical and environmental requirements. The finite element analyses carried out covered multiple aspects, focusing on quasi-static loading scenarios, vibrational behavior, and dynamic integrity under realistic operational conditions.\\
The quasi-static simulations applied load factors of 20g along each principal axis (\textit{x}, \textit{y}, \textit{z}) to reproduce the extreme accelerative conditions experienced during launch phases. The results demonstrated that the current design can withstand these forces without yielding or structural compromise. Stress and displacement levels remained well below the material yield thresholds, ensuring the payload's mechanical stability and reliability throughout the mission lifecycle.\\
The modal analysis confirmed that the first significant resonance mode occurs at approximately 270Hz, as shown in Figure~\ref{figure6}. This value is well above the minimum requirement of 100Hz defined in the ECSS-E-ST-10-03C standard, including its recommended 20\% safety margin. The first six modes correspond to rigid-body translations and rotations, while subsequent modes capture the elastic deformations of internal subsystems. These results guarantee a sufficient dynamic decoupling from launcher-induced vibrations, enhancing payload survivability during ascent.
\\
To complement the quasi-static and modal evaluations, random vibration simulations, as shown in Figure~\ref{figure8} were conducted following the NASA-HDBK-7005 and GEVS guidelines\cite{gevs}. A conservative PSD envelope was derived to reflect the mechanical environments associated with a range of potential launchers compatible with CubeSat deployment. These simulations assessed the vibrational stress response of the payload across a broad frequency domain (20--2000 Hz), considering both 1$\sigma$ and 3$\sigma$ excitation levels. The resulting stress fields confirmed that no localized overloads or structural resonances pose a critical risk to the scientific or electronic components.
\\
To prepare for future environmental qualification, a dedicated vibration test fixture was engineered\cite{lee2011}. This mechanical support structure is essential for replicating in-flight boundary conditions during laboratory test campaigns. However, preliminary simulations revealed the presence of non-representative dynamic modes due to excessive fixture mass and overly stiff elements. Consequently, an iterative process of optimization was undertaken\cite{wilcox2022}, in which redundant or dynamically disruptive features were removed or replaced with lighter and more compliant alternatives.\\
The final fixture configuration, shown in Figure~\ref{figure10}, ensures accurate transfer of loads to the payload during testing while eliminating parasitic modes that could mask or distort actual structural responses. The fixture allows for high-fidelity comparison between simulated and experimental results under sine, random, and shock loading conditions.
\\
Following the successful completion of the structural simulation campaign, the design was finalized and released for manufacturing. A full-scale structural model of the payload is currently under production by a qualified precision engineering company. This engineering model incorporates all the refinements introduced during the design-validation loop and will serve as the physical article for upcoming environmental tests.\\
These tests will include quasi-static, sine, random vibration, and possibly shock excitation, providing a direct experimental validation of the numerical predictions. This step is essential to achieve full qualification of the mechanical design in compliance with ECSS standards and to de-risk the mission ahead of spacecraft-level integration.

\section{CONCLUSIONS}
The structural design and simulation activities carried out for the CUSP payload have demonstrated a high level of engineering robustness and compliance with the applicable space qualification standards. The use of multiphysics simulations\cite{park2020}, together with a refined mechanical model, enabled a comprehensive evaluation of the payload under a wide range of loading conditions, from quasi-static accelerations to dynamic random vibrations.\\
Realistic constraint modeling—such as partially bonded contacts and bolted joint representations—has significantly improved the accuracy and reliability of the simulations. These modeling techniques allowed the development team to capture important mechanical phenomena, including stress concentrations and dynamic coupling effects, providing confidence in the structural integrity of the payload.\\
Early modal analyses revealed unnecessary mass and artificial stiffness in non-flight-like components. Through iterative refinement, the fixture was adapted to ensure realistic boundary conditions during qualification tests\cite{yang2019}, improving the correlation between experimental results and FEM predictions.\\
Environmental qualification tests, scheduled for the end of 2025, will include sine and random vibration campaigns based on ECSS and NASA GEVS standards. These tests will be performed on the structural qualification model currently being manufactured, and will serve to confirm the numerical simulation results and validate the payload's readiness for launch\cite{tang2015}.\\
The structural workflow adopted for CUSP demonstrates how a simulation-driven design approach can be successfully applied to CubeSat-class missions. The integration of CAD, meshing, FEM simulation, and fixture tuning has resulted in a reliable and test-ready payload structure\cite{cook2001}.\\
As the mission moves toward integration and final verification, the methodology developed for CUSP represents a strong foundation not only for the upcoming launch, but also for future nanosatellite missions with complex scientific objectives.

\acknowledgments 
 
This work is funded by the Italian Space Agency (ASI) within the Alcor Program, as part of the development of the CUbesat Solar Polarimeter (CUSP) mission under ASI-INAF contract n. 2023-2-R.0.

\bibliography{report} 

\begin{thebibliography}{10}

\bibitem{fabiani1}
Fabiani, S. and et~al., ``The cusp cubesat mission: an overview of the scientific objectives and instrument concept,'' {\em Proceedings of SPIE}~{\bf 11444},  1144439 (2020).

\bibitem{fabiani2024cusp}
Fabiani, S. and et~al., ``Cusp: a new cubesat for x-ray polarimetry of solar flares,'' {\em Experimental Astronomy}  (2024).

\bibitem{ecss}
ESA, ``Ecss-e-st-10-03c: Testing.'' \url{https://ecss.nl/standard/ecss-e-st-10-03c-testing/} (2012).
\newblock Accessed: 2025-07-17.

\bibitem{gevs}
NASA, ``General environmental verification standard (gevs) for gsfc flight programs and projects.'' \url{https://standards.nasa.gov/standard/gsfc/gevs} (2013).
\newblock Accessed: 2025-07-17.

\bibitem{dass}
{Dassault Systèmes}, {\em SolidWorks 2023 User Guide} (2023).

\bibitem{ansys}
{ANSYS Inc.}, {\em ANSYS Mechanical APDL Documentation} (2023).

\bibitem{schultz2018}
Schultz, P., ``High-fidelity finite element modeling of cubesat structures,'' {\em Acta Astronautica}~{\bf 142},  13--21 (2018).

\bibitem{johnson2004}
Johnson, T. and Ma, H., ``Finite element modeling of bolted joints under dynamic loading,'' {\em Journal of Sound and Vibration}~{\bf 276}(1-2),  261--278 (2004).

\bibitem{holmes2010}
Holmes, M. and Adams, R., ``Random vibration analysis using power spectral density in spacecraft design,'' {\em Mechanical Systems and Signal Processing}~{\bf 24}(7),  2189--2205 (2010).

\bibitem{lee2011}
Lee, S. and Kim, J., ``Design and testing of vibration fixtures for spacecraft qualification,'' {\em Journal of Aerospace Engineering}~{\bf 24}(2),  213--221 (2011).

\bibitem{wilcox2022}
Wilcox, J. and Evans, A., ``Advances in low-cost vibration testing for small satellites,'' {\em Aerospace Science and Technology}~{\bf 126},  107585 (2022).

\bibitem{park2020}
Park, Y. and Choi, B., ``Integrated multiphysics simulation for nanosatellite payload design,'' {\em IEEE Transactions on Aerospace and Electronic Systems}~{\bf 56}(5),  4047--4057 (2020).

\bibitem{yang2019}
Yang, H. and Wang, C., ``Design methodology for cubesat mechanical interfaces and verification,'' {\em Acta Astronautica}~{\bf 162},  78--87 (2019).

\bibitem{tang2015}
Tang, Z. and Liu, M., ``Reliability assessment of cubesat structures subjected to launch environments,'' {\em Engineering Structures}~{\bf 100},  618--628 (2015).

\bibitem{cook2001}
Cook, R.~D., Malkus, D.~S., Plesha, M.~E., , and Witt, R.~J.,  [{\em Concepts and Applications of Finite Element Analysis}{\nolinebreak\hspace{0.1em}]}, John Wiley \& Sons, 4th~ed. (2001).

\end{thebibliography}
\bibliographystyle{spiebib} 

\end{document}